\documentclass{achemso}

\usepackage[version=3]{mhchem}
\usepackage{notes2bib}
\usepackage{natmove}
\usepackage{graphicx}
\usepackage{dcolumn}
\usepackage{pifont}
\usepackage{bm}
\usepackage{multirow}
\usepackage{amsmath}
\usepackage{hyperref}
\usepackage{booktabs}
\usepackage{float}
\usepackage{txfonts}

\usepackage[dvips]{color}
\definecolor{darkgreen}{rgb}{0,.6,0}

\author{Gum-Chol Ri}
\author{Chol-Jun Yu}
\email{ryongnam14@yahoo.com}
\author{Jin-Song Kim}
\author{Song-Nam Hong}
\author{Un-Gi Jong}
\author{Mun-Hyok Ri}
\affiliation{Computational Materials Design (CMD), Faculty of Materials Science, Kim Il Sung University, Ryongnam-Dong, Taesong District, Pyongyang, DPR Korea}

\title{Insertion of alkali atoms into graphite enhanced by cointercalation with alkylamine: mechanism from density functional theory}

\begin{document}

\begin{abstract}
Using density functional theory calculations, we have investigated the structural, energetic, and electronic properties of ternary graphite intercalation compounds (GICs) containing alkali atoms (AM) and normal alkylamine molecules (nC$x$), denoted as AM-nC$x$-GICs (AM=Li, Na, K; $x$=1, 2, 3, 4). The orthorhombic unit cells have been used to build the models for crystalline stage-I AM-nC$x$-GICs. By performing the variable cell relaxations and the analysis of results, we have found that with the increase in the atomic number of alkali atoms the layer separations decreases in contrast to AM-GICs, while the bond lengths of alkali atoms with graphene layer and nitrogen atom of alkylamine decreases. The formation and interlayer binding energies of AM-nC3-GICs have been calculated, indicating the increase in stability from Li to K. The calculated energy barriers for migration of alkali atoms suggest that alkali cation with larger ionic radius diffuses in graphite more smoothly, being similar to AM-GICs. The analysis of density of states, electronic density differences, and atomic populations illustrates a mechanism how the insertion of especially Na among alkali atoms into graphite with first stage can be made easy by cointercalation with alkylamine; more extent of electronic charge transfer is occurred from more electropositive alkali atom to carbon ring of graphene layer, while alkylamine molecules interact strongly with graphene layer through the hybridization of valence electron orbitals.
\end{abstract}

%\begin{keyword}
%Graphite intercalation compound \sep Alkali metal \sep Alkylamine \sep Density functional theory \sep Cointercalation
%\end{keyword}
%\maketitle

\section{\label{sec:intro}Introduction}
Graphite is composed of honeycomb carbon layer planes (so-called graphene layers) weekly bound through van der Waals (vdW) interaction, and thus incorporates readily a vast range of guest dopants between layers to form graphite intercalation compounds (GICs)~\cite{Dresselhaus}. GICs have interesting physico-chemical properties such as broad range of electrical conductivity behavior from insulating to superconducting, high lubricity, and high chemical reactivity, which differ entirely from the parents, opening a plethora of industrial applications~\cite{Csanyi,Matsumoto}.

Alkali atoms are probably the most widely studied intercalants that can be inserted with ease except Na into graphite host material. They form electron donor compounds by nature and some exhibit superconductivity~\cite{Csanyi,Hannay,Belash}, for example, the C$_8$K system with $T_c\simeq 0.14$~K~\cite{Hannay}. Recently, alkali metal GICs (AM-GICs) have been renewed as anode materials for energy storage applications such as Li-ion batteries (LIBs)~\cite{Buldum,Tarascon,Zhu10,Goodenough} and Na-ion batteries (NIBs)~\cite{Nobuhara,Slater,Yabuuchi12,Yabuuchi14,Nose}. Since sodium is much more abundant than lithium on the earth's crust, NIBs are more cost-effective than LIBs, provided that high performance active electrode materials for NIBs could be discovered. However, there has been a lot of works to report a severe problem hindering the development of NIBs with high efficiency; first-principles calculations of formation energies and energy barriers of AM-GICs (AM: Li, Na, K)~\cite{Nobuhara,Okamoto} indicated that Na hardly intercalates into graphite due to the energetic instability of Na-GICs in contrast to Li- and K-GICs, and the larger radius ions more smoothly diffuse in graphite. The lattice mismatch of graphite layers and of solid Na has been proposed as another potential reason for absence of significant Na insertion into graphite\cite{Dresselhaus79}. Failure in formation of low stage Na-GICs was also confirmed by experiment~\cite{Stevens}. As a possible solution to this problem, either expanded graphite with an enlarged interlayer distance of 4.3 \AA, which is prepared through a process of oxidation and partial reduction~\cite{Wen14}, or nano-sized graphite (so-called amorphous carbon)~\cite{Lotfabad,Legrain,Tsai} has been suggested as a potential host for Na-GICs.

Cointercalation process to form ternary graphite intercalation compounds (t-GICs) is recognized as a promising method for the insertion of materials that do not readily intercalate by themselves. By this type of process, Na can intercalate in low stage compounds with another intercalants of small molecules such as ammonia or ammonium ion~\cite{Walters,Reghai} and with some electrolytes~\cite{Matsumoto,Kim14}. It is a great advance in the NIB development that Jache and Adelhelm reported a novel NIB with low overpotentials and stable capacities of around 100 mAh/g for more than 1000 cycles, which uses a stage-I t-GIC as anode with an estimated stoichiometry of Na(diglyme)$_2$C$_{20}$ prepared by using a diglyme-based electrolyte~\cite{Jache}. In recent years, Lerner's group~\cite{Maluangnont1,Maluangnont2,Maluangnont3} synthesized a series of t-GICs containing alkali metals and alkylamines, and studied the structures, compositions, and reaction chemistries by using powder X-ray diffraction. In fact, alkylamines or alkylammonium ions intercalation compounds have a high chemical reactivity as they are applied to detoxification, photo- and electro-functional devices, and catalysts~\cite{Ogawa}.

It can be thought that, in addition to providing a method easy for hard intercalants, cointercalation process has a potential effect on the chemical reactivity, either enhancing the reactivity with donor-donor and acceptor-acceptor cointercalation or reducing the reactivity with donor-acceptor cointercalation. The former effect can be positive for improving the performance of alkali metal ion batteries but probably negative for the stability of battery material with its environment. However, it is not clear yet what effect the cointercalation process with alkali metals and such organic molecules as alkylamines has on the performance and stability of materials.

In this paper, we apply the state-of-the-art {\it ab initio} density functional theory (DFT) to the t-GICs containing alkali metals (AM: Li, Na, K) and alkylamines (alkyl: methyl, ethyl, propyl, butyl) to evaluate the tendency of structural stability, energetics, electronic structure, and atomic charge population. We obtain the fully optimized crystal structures of these materials with and without the inclusion of vdW dispersion correction to the total energy. We pay attention to the analysis of electronic properties of these GICs to find the charge transfer in detail. In the following, we organize this work as Section~\ref{sec:theoretics} for description of computational method and models, Section~\ref{sec:result} for results and discussion, and finally Section~\ref{sec:con} for conclusions.

\section{\label{sec:theoretics}Computational method and model}
\begin{figure}[t]
\begin{center}
\includegraphics[clip=true,scale=0.18]{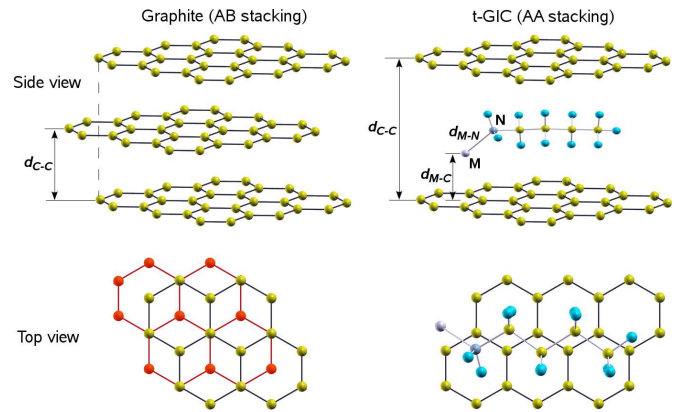}
\caption{\label{fig1}(Color online) Illustration of natural hexagonal graphite with $AB$ stacking and ternary GICs containing linear alkylamine and alkali atoms between $AA$-stacked graphene layers with expanded layer separation $d_{C-C}$. Yellow-colored spheres represent for carbon atoms, grey-colored for alkali atoms, indigo-colored for nitrogen atom, and blue-colored for hydrogen atoms.}
\end{center}
\end{figure}

We begin with crystalline lattice modeling of t-GICs containing alkali atoms (AMs) and alkylamine molecules with a certain number of carbon atoms, where selected AMs are Li, Na, and K, and alkyl groups are methyl, ethyl, propyl, and butyl. Note that in the experimental synthesis~\cite{Maluangnont1} methyl and ethyl were not attempted. With respect to the alkylamine molecules, we adopt only normal (linear) structure for the sake of simplicity, although other conformations like {\it iso}- or {\it sec}- (branched) structures are not ruled out in the experiment~\cite{Maluangnont1}. While natural graphite has an $AB$ stacking with the graphite sheets shifted relative to each other, the cointercalants composed of linear alkylamine and AMs in t-GICs are placed between $AA$-stacked graphene layers, whose layer separation $d_{C-C}$ is expanded compared with that of graphite, as illustrated in Figure~\ref{fig1}.

Here, the normal alkylamine molecules are denoted as nC$x$ (x=1, 2, 3, 4), and, considering that an appropriate number of carbon atoms in the graphene layer is estimated to be 24 for normal propylamine in the experiment~\cite{Maluangnont1}, the t-GICs studied in this work can be notated by AM-nC$x$-GICs or stoichiometrically M(nC$x$)C$_{24}$, which are all stage-I t-GICs. To model these GICs, we use orthorhombic unit cells for crystalline lattices and determine the lattice constants by performing crystalline lattice optimization while allowing atomic relaxation.

The calculations were carried out within the framework of density functional theory (DFT). The geometries obtained by modeling were fully relaxed with pseudopotential plane wave method as implemented in Quantum ESPRESSO package~\cite{pwscf}. While ultrasoft pseudopotentials (USPP) provided in the package were used for C, N, H, Li, and Na~\cite{uspp}, we generated our own USPP for potassium by using the Atomic code accompanied with the package~\cite{pwscf}. The kinetic energy cutoff for plane wave expansion is set to be 40 Ry and $k$-points for the first Brillouin zone integration are (2$\times$4$\times$6). These calculation parameters guarantee an accuracy of the total energy as 5 meV per cell. In the structural optimization, the atoms were relaxed until the forces on atoms converge into 0.02 eV/\AA.

The accuracy of DFT predictions depends mostly on the approximation to exchange-correlation (XC). There are two contrastive functionals widely used in DFT calculations, constructed within generalized gradient approximation (GGA) and local density approximation (LDA); GGA functionals in general underestimate the binding energy, whereas LDA usually overbinds. It is accepted that, for graphite, GGA fails significantly to produce any binding between the layers, but LDA produces a good prediction of the equilibrium interplanar spacing. Moreover, there has been a lot of reports that emphasize a great importance of vdW correction for a good prediction of materials properties and processes of graphite-related materials (for example, see Refs.~\cite{yucj14,Ziambaras}). To form a judgment on the choice of functional, we have tested different XC functionals including Perdew-Burke-Ernzerhof (PBE)~\cite{pbe} favor and PBE revised version for solid (PBEsol)~\cite{PBEsol} within GGA, Perdew-Zunger (PZ)~\cite{PZlda} formula within LDA, and PBE plus vdW correction (vdW-DF4)~\cite{vdwDF4}.

The stability of AM-nC$x$-GICs can be estimated by calculating the formation energy for the reaction, $[\text{M}+\text{nC}x+6\text{C}_4\rightarrow \text{M(nC}x\text{)C}_{24}]$, using the following equation,
\begin{equation}
\label{eq:form}
E_{\text{f}}=E_{\text{M(nC}x\text{)C}_{24}}-(E_{\text{M}}+E_{\text{nC}x}+6E_{\text{C}_4}),
\end{equation}
where $E_{\text{M(nC}x\text{)C}_{24}}$, $E_{\text{nC}x}$, and $E_{\text{C}_4}$ are total energies of orthorhombic unit cell of AM-nC$x$-GIC crystal, supercell containing an isolated alkylamine molecule with adjacent molecular distance of about 12 \AA, and hexagonal unit cell of graphite composed of 4 carbon atoms, and $E_{\text{M}}$ is total energy per atom of $bcc$ unit cell for alkali metal, respectively. Negative values of $E_f$ indicate therefore that t-GIC formation is thermodynamically favorable, while positive values indicate unfavorableness of the formation.

As supplement, we applied SIESTA code, which employs norm-conserving pseudopotentials and atomic orbital basis sets~\cite{SIESTA}, to make an analysis of atomic charge population with Mulliken, Hirshfeld, and Voronoi methods~\cite{Hirshfeld}. Standard basis sets of double zeta plus polarization (DZP) and Troullier-Martins type pseudopotentials~\cite{TMpseudo} generated by ourselves using PBEsol functional were used. Mesh cutoff that defines the equivalent plane wave cutoff for the real space grid, and $k$-grid cutoff which determines the fineness of the reciprocal space grid used for Brillouin zone sampling were set to be 200 Ry and 10 \AA, respectively.

\section{\label{sec:result}Results and discussion}

\subsection{\label{subsec:structure}Structure}
We first describe the effect of different XC functionals on structural optimization. For each of selected functionals, $bcc$ unit cells of alkali metals and hexagonal graphite unit cell were fully optimized with an automatic way of PWscf code (variable cell relaxation). Supercells were constructed to model isolated alkylamine molecules using orthorhombic lattices with lattice constants of $a=23$ \AA~and $b=c=15$ \AA, which are enough wide to inhibit artificial interaction between the molecule and its periodic images by providing the intermolecular distance of about 12 \AA. The alkylamine molecule with linear chain structure is aligned along the $a$ lattice direction in the supercell, and all the atomic coordinates of the molecule were fully relaxed. The orthorhombic lattices for t-GICs were built, starting from the hexagonal unit cell of graphite with experimental lattice constants of $a=2.461$ \AA~and $c=6.705$ \AA~($d_{\text{C-C}}=c/2=3.353$ \AA)~\cite{boettger} and using $(\sqrt{3}a\times6a\times1c)$ unit cell. Then, the relaxed alkylamine molecule was inserted into a space between $AA$-stacked graphene layers formed in orthorhombic unit cells of AM-nC$x$-GICs (see Figure~\ref{fig1}), and all the atomic coordinates were relaxed again, allowing only lattice constants to be changed while fixing the lattice angles.

With respect to the structural optimization of graphite unit cell, the experimental in-plane lattice constant, $a$=2.461 \AA~\cite{boettger}, was well reproduced by all XC functionals, but the inter-planar lattice constant is severely dependent on the choice of XC functionals; 6.603 \AA~(PZ), 6.563 \AA~(vdW-DF4), 7.091 \AA~(PBEsol), and 7.951 \AA~(PBE). Therefore, it is again confirmed that LDA (PZ) yields the value closest to the experiment (6.705 \AA) with a relative error of $\sim1.5\%$, whereas GGA (PBE) produces unrealistic value with a relative error over $18\%$. We should note that though the value found from vdW-DF4 is in good agreement with the experiment ($\sim2.1\%$) as well, PBEsol dose not give really bad result ($\sim5.8\%$).

\begin{figure}[!t]
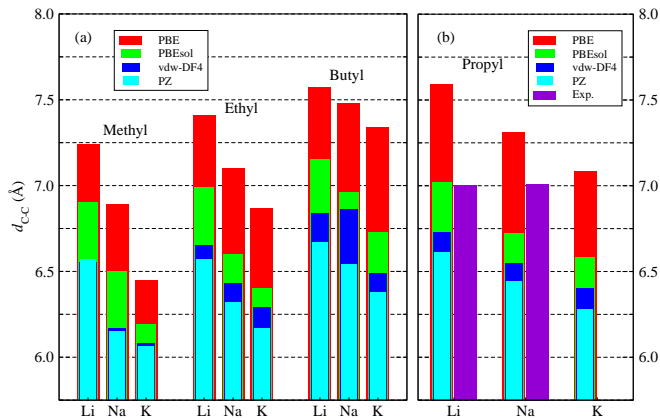

\begin{center}
\includegraphics[clip=true,scale=0.34]{fig2a.eps}
\includegraphics[clip=true,scale=0.34]{fig2b.eps}
\caption{\label{fig:dcc}(Color online) Graphene layer separation in AM-nC$x$-GICs, calculated with different XC functionals; (a) for methyl, ethyl, and butyl, and (b) for propyl with the experimental values for Li -- Li(nC3)$_{0.8}$C$_{16}$ -- and for Na -- Na(nC3)$_{0.7}$C$_{16}$~\cite{Maluangnont1}. }
\end{center}
\end{figure}

In Figure~\ref{fig:dcc}, we presented the layer separation $d_{\text{C-C}}$ of AM-nC$x$-GICs obtained by structural optimization. At a glance, it is found that PBE gives the largest estimation, while PZ the smallest, and the values by PBEsol and vdW-DF4 are between them. Figure~\ref{fig:dcc} (b) shows that PBEsol produces more reliable value rather than vdW-DF4, comparing with the experimental data~\cite{Maluangnont1}. Note that the experimental values for Li and Na are for Li(nC3)$_{0.8}$C$_{16}$ and Na(nC3)$_{0.7}$C$_{16}$, corresponding to Li(nC3)C$_{20}$ and Na(nC3)C$_{23}$ in our notation, respectively. We thought that the coincidence by PBEsol is not surprising because it can be expected that in stage-I t-GICs the interaction between graphene layers could be no more week vdW due to the presence of intercalants. Together with the above argument for graphite, therefore, we can make a suggestion that PBEsol can reliably predict material properties of stage-I AM-nC$x$-GICs.

With all tested XC functionals, the layer separation increases when increasing the number of carbon atoms of alkylamines for all alkali metals, but decreases monotonically with the increase in the atomic number of alkali atoms for all alkylamines. Note that a long plateau in the layer separation is observed from nC6 to nC14 in the experiment~\cite{Maluangnont1}, which is contrasted with a gradual increase usually observed on intercalation of other layered hosts, as well as on our computational results. What we want to stress here is the variation of $d_{\text{C-C}}$ with the increase in the atomic number of alkali atoms, comparing between AM-nC$x$-GICs and AM-GICs; it decreases in AM-nC$x$-GICs versus increases in AM-GICs~\cite{Nobuhara}. From such contrary observation, we can anticipate an important role of alkylamine molecule, facilitating the insertion of alkali atoms into graphite, as will be discussed below.

\begin{table}[!b]
\small
\begin{center}
\caption{\label{tab:bondlenths}Bond lengths of alkali atom with carbon atom of bottom graphene layer ($d_{\text{M-C}}$) and with nitrogen atom of alkylamine molecule ($d_{\text{M-N}}$) in angstrom unit, estimated with different XC functionals.}
\begin{tabular}{l@{}c@{\hspace{5pt}}c@{\hspace{5pt}}c@{\hspace{5pt}}|c@{\hspace{5pt}}c@{\hspace{5pt}}|c@{\hspace{5pt}}c@{\hspace{5pt}}|c@{\hspace{5pt}}c}
\hline
      &       & \multicolumn{2}{c|}{PBE} & \multicolumn{2}{c|}{PBEsol} & \multicolumn{2}{c|}{vdW-DF4}& \multicolumn{2}{c}{PZ} \\
\cline{3-10} 
Alkyl & Metal & $d_{\text{M-C}}$ & $d_{\text{M-N}}$ & $d_{\text{M-C}}$ & $d_{\text{M-N}}$ & $d_{\text{M-C}}$ & $d_{\text{M-N}}$ & $d_{\text{M-C}}$ & $d_{\text{M-N}}$ \\
\hline
       & Li   & 2.00 & 2.11 & 1.98 & 2.07 & 2.06 & 2.07 & 1.94 & 2.03 \\
Methyl & Na   & 2.59 & 2.40 & 2.32 & 2.34 & 2.72 & 2.40 & 2.34 & 2.31 \\
       & K    & 3.20 & 2.79 & 3.11 & 2.75 & 2.99 & 2.81 & 3.00 & 2.72 \\
\hline
       & Li   & 2.21 & 2.09 & 2.17 & 2.07 & 2.26 & 2.08 & 1.99 & 2.04 \\
Ethyl  & Na   & 2.44 & 2.33 & 2.41 & 2.35 & 2.81 & 2.59 & 2.35 & 2.11 \\
       & K    & 2.82 & 2.79 & 3.21 & 2.75 & 3.19 & 2.80 & 3.14 & 2.70 \\
\hline
       & Li   & 2.23 & 2.08 & 2.08 & 2.06 & 2.30 & 2.08 & 2.06 & 2.03 \\
Propyl & Na   & 2.42 & 2.38 & 2.37 & 2.34 & 2.70 & 2.41 & 2.31 & 2.30 \\
       & K    & 2.88 & 2.80 & 2.89 & 2.75 & 3.11 & 2.79 & 2.92 & 2.70 \\
\hline
       & Li   & 2.22 & 2.10 & 2.11 & 2.07 & 2.13 & 2.08 & 2.08 & 2.04 \\
Butyl  & Na   & 2.44 & 2.38 & 2.37 & 2.36 & 2.36 & 2.40 & 2.40 & 2.31 \\
       & K    & 2.82 & 2.79 & 2.83 & 2.73 & 3.19 & 2.80 & 2.86 & 2.70 \\
\hline
\end{tabular}
\normalsize
\end{center}
\end{table}
We then discuss bonding characteristics of alkali atom with carbon atom of bottom graphene layer and nitrogen atom of alkylamine, i.e., $d_{\text{M-C}}$ and $d_{\text{M-N}}$ as illustrated in Figure~\ref{fig1}. The results are listed in Table~\ref{tab:bondlenths}. We can recognize that with the variation of the number of carbon atoms of alkylamine, there is no clear rule in both $d_{\text{M-C}}$ and $d_{\text{M-N}}$ variation for all alkali metals, even being almost unchangeable for the case of $d_{\text{M-N}}$. This indicates the similar effect of different alkylamine molecules on alkali atoms. On the other hand, with the increase in the atomic number of alkali atoms, both $d_{\text{M-C}}$ and $d_{\text{M-N}}$ increase gradually for all alkylamines, in contrast to the variation of $d_{\text{C-C}}$ as discussed above.
\begin{figure*}[!t]
\begin{center}
\includegraphics[clip=true,scale=0.25]{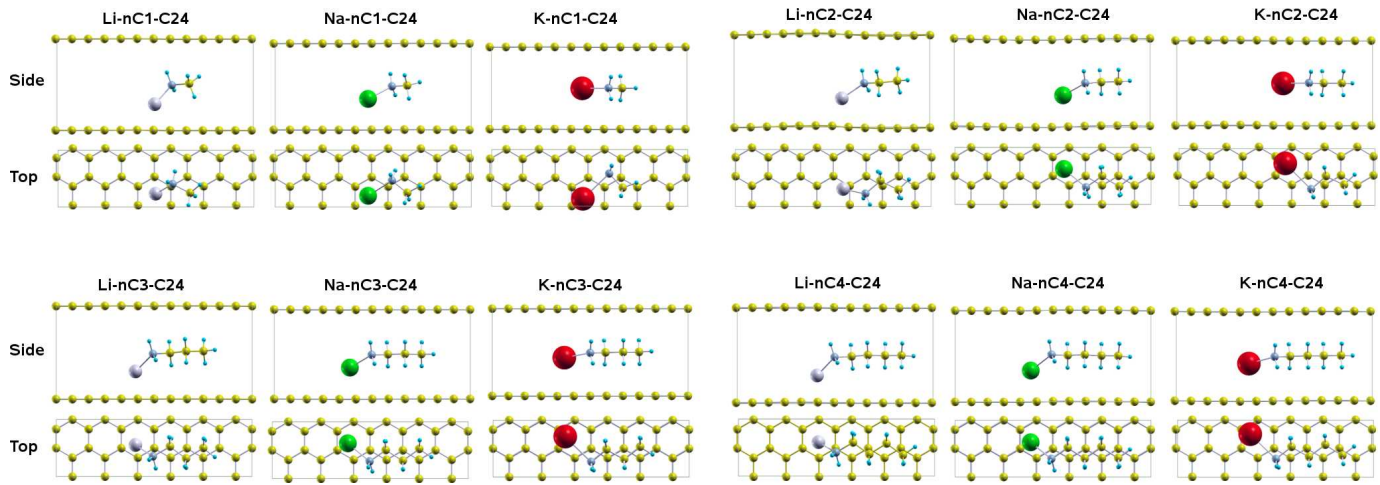}
\caption{\label{fig-opt}(Color online) Structure of ternary graphite intercalation compound AM-nC$x$-GICs with orthorhombic crystalline lattices, optimized using PBEsol functional. With the increase of atomic number of alkali atoms, the layer separation decreases, while bond lengths of alkali atoms with bottom graphene layer and with nitrogen atom increase.}
\end{center}
\end{figure*}

To more intuitively show the variations of $d_{\text{C-C}}$, $d_{\text{M-C}}$ and $d_{\text{M-N}}$, we depicted the crystalline structures of t-GICs optimized with PBEsol in Figure~\ref{fig-opt}. When increasing the atomic number of alkali atoms, the increase in $d_{\text{M-C}}$ or $d_{\text{M-N}}$ indicates stronger interaction of less electronegative (or more electropositive) alkali atom with graphene layer or nitrogen atom, while the decrease in $d_{\text{C-C}}$ indicates stronger binding of graphene layers through the cointercalants composed of less electronegative alkali atom and alkylamine molecule. Therefore it can be suggested that alkylamine molecule has a great impact, which intercepts more expanding effect of alkali metal cation with larger ionic radius and conversely contracts the layer separation.

\subsection{\label{subsec:energetics}Energetics}
\begin{figure}[b!]
\begin{center}
\includegraphics[clip=true,scale=0.5]{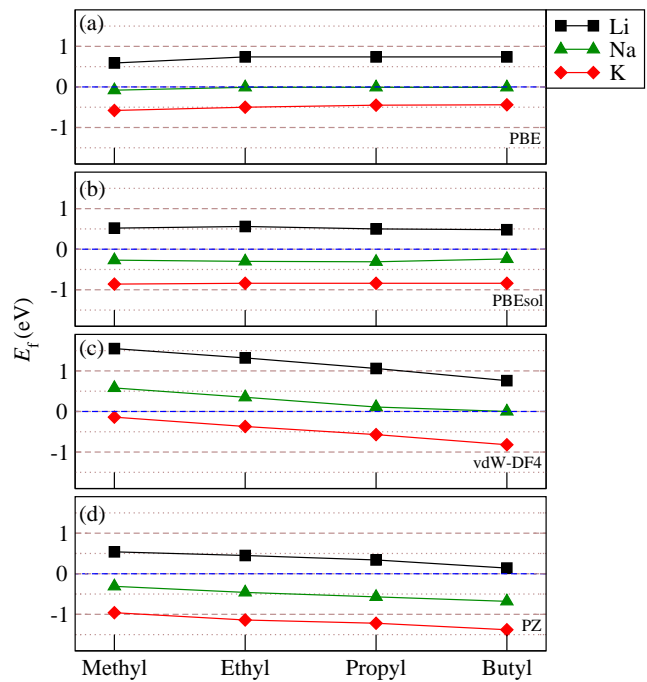} \\
\caption{\label{fig:forme}(Color online) Calculated formation energies of AM-nC$x$-GICs with different XC functionals; (a) for PBE, (b) for PBEsol, (c) for vdW-DF4, and (d) for PZ functional.}
\end{center}
\end{figure}

The structural stability can be estimated by energetics as well. Firstly, we consider the formation energies ($E_{\text{f}}$) of AM-nC$x$-GICs as plotted in Figure~\ref{fig:forme}. 
With all different XC functionals, it turns out that Li-nC$x$-GICs (x=1, 2, 3, 4) are energetically unstable from their positive formation energies, while K-nC$x$-GICs being stable from their negative ones. The stability of Na-nC$x$-GICs is dependent on the choice of XC functionals; they are expected to be energetically stable with PZ and with PBEsol that gives the most reliable layer separation as considered above, whereas with PBE and vdW-DF4 they are predicted to be unstable. If we rely on PBEsol functional, Na-nC$x$-GICs should be energetically stable. We note that the variation tendency of formation energies $E_{\text{f}}$ when increasing the atomic number of alkali atoms is similar to that of layer separations $d_{\text{C-C}}$, providing the same assertion of stronger interaction between more electropositive alkali atom (or alkali metal cation with larger ionic radius) and alkylamine molecule as well as graphene layer.

When comparing with those of AM-GICs calculated using PBE in the previous DFT work~\cite{Nobuhara}, K-GICs have the lowest formation energies among three alkali metal GICs as well, and Li-GICs are always stable for all considered stages. For Na-GICs, however, only high stage GICs (NaC$_{16}$ and NaC$_{12}$) are stable but low stage GICs (NaC$_8$ and NaC$_6$) are unstable. Therefore, as forecasted in the introduction of this paper, it may be deduced that cointercalation of alkali atoms with alkylamines provide a way for forming Na GICs with first stage.

\begin{figure}[t!]
\begin{center}
\includegraphics[clip=true,scale=0.095]{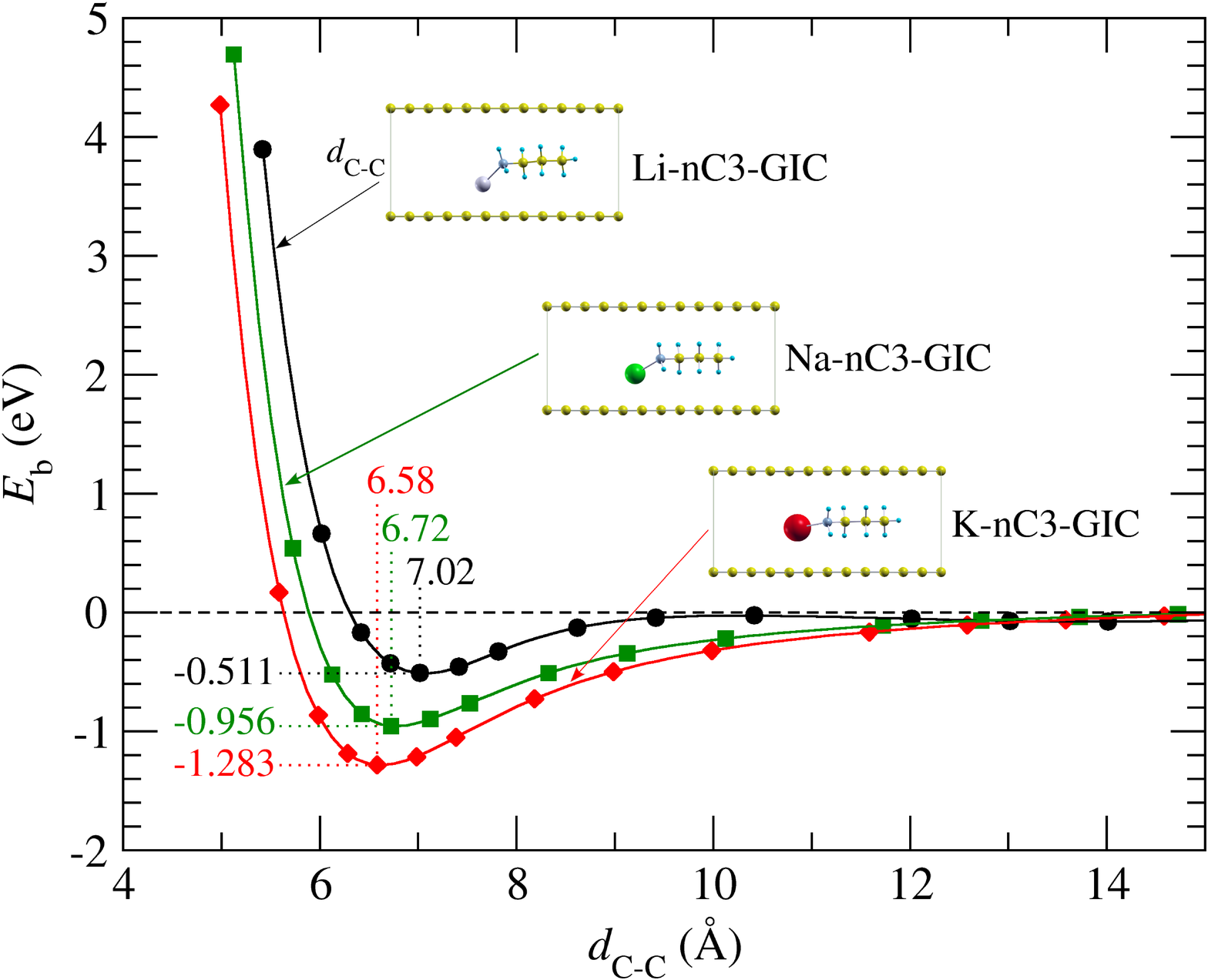} \\
\caption{\label{fig:binde}(Color online) Interlayer binding energies and corresponding layer separations of AM-nC3-GICs, calculated using PBEsol XC functional.}
\end{center}
\end{figure}

The chemical bonding properties of AM-nC$x$-GICs can be assessed more quantitatively by interlayer binding energies $E_{\text{b}}$, which were calculated as the total energy difference between the lowest energetic state and the state with infinite layer separation, when expanding it gradually. Typically, the results of AM-nC3-GICs calculated with PBEsol functional are plotted in Figure~\ref{fig:binde}. It is shown that the interlayer binding energies are estimated to be -0.511 eV, -0.956 eV, and -1.283 eV with the layer separations of 7.02 \AA, 6.72 \AA, and 6.58 \AA~for Li-, Na-, and K-nC3-GICs, respectively. The results indicate again that the order of binding strengths is like Li$<$Na$<$K. Note that the layer separations are in agreement with those obtained by structural optimization.

\begin{figure}[t!]
\begin{center}
\includegraphics[clip=true,scale=0.095]{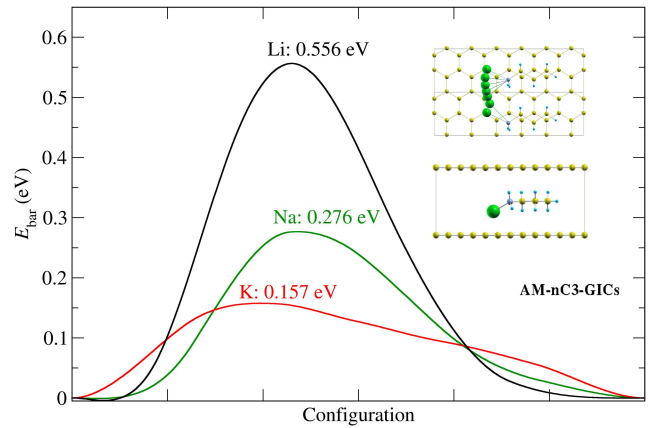} \\
\caption{\label{fig:bar}(Color online) Energy barriers for migrations of alkali metal cations in AM-nC3-GICs, calculated by using NEB method with PBEsol XC functional. Seven NEB images were used to construct a path shown in inset figure.}
\end{center}
\end{figure}

We then calculated the energy barriers ($E_{\text{bar}}$) for migration of the alkali metal cations from the site determined by atomic relaxation to the adjacent image site in AM-nC3-GICs, using PBEsol XC functional. The calculations were carried out by using nudged elastic band (NEB) method implemented in PWneb code. We have used 7 NEB image configurations to discretize the path. As shown in Figure~\ref{fig:bar}, the calculated energy barriers are 0.556 eV, 0.276 eV, and 0.157 eV for Li-, Na-, and K-nC3-GICs, respectively; that is, the cation with larger ionic radius has smaller energy barrier, indicating more smoothly diffuse in graphite. Interestingly, the tendency of energy barriers for AM-nC3-GICs is the same to that of AM-GICs~\cite{Nobuhara}, in contrast to the cases of formation and binding energies. Here, we want to emphasize that sodium alkylamine GICs have stronger binding and more smoothly diffuse in graphite than lithium alkylamine GICs, which might be positive aspects for NIBs.

\subsection{\label{subsec:charge}Electronic properties}
\begin{figure}[t!]
\begin{center}
\includegraphics[clip=true,scale=0.47]{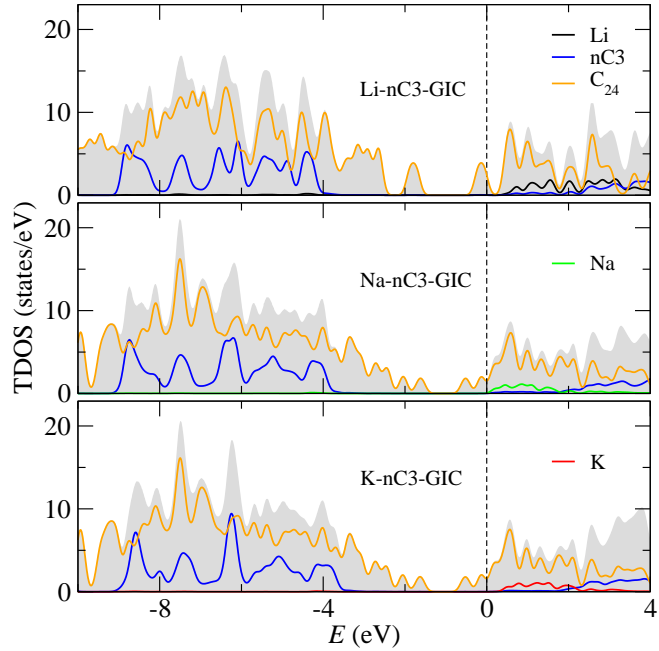} \\
\caption{\label{fig:dos}(Color online) Total density of states (TDOS) for AM-nC3-GICs, calculated with PBEsol XC functional. TDOS (grey background) is decomposed into contributions from alkali atom (black, green, and red line for Li, Na, and K, respectively), alkylamine molecule (blue line) and graphene layer (orange line). The dashed vertical line indicates the Fermi level placed at 0 eV.}
\end{center}
\end{figure}

To further understand the material properties of AM-nC$x$-GICs and seek out mechanism behind them, we consider electronic properties such as density of states (DOS), electronic density differences, and atomic charge populations. They can provide us an invaluable insight how charge transferring occurs in the event of GIC formation. In Figure~\ref{fig:dos}, we show the total density of states (TDOS) with atomic resolved TDOS of AM-nC3-GICs calculated with PBEsol functional. Here, for all the DOS plots, the Fermi level is set to be 0 eV on the energy axis. Apparently, all the AM-nC3-GICs are metallic, since there exist valence electron states at the Fermi level, originated from delocalized $p_z$ electron of carbon atom forming $\pi$ bond in graphene layer. We can find that TDOS of carbon atoms of graphene layer overlaps with that of propylamine below Fermi level, whereas with those of alkali atoms over Fermi level. Therefore, it can be stated that alkylamine molecules interact rather strongly with graphene layers through the hybridization between their electron states, while alkali atoms release electrons, becoming cations.

\begin{figure}[t!]
\begin{center}
\includegraphics[clip=true,scale=0.14]{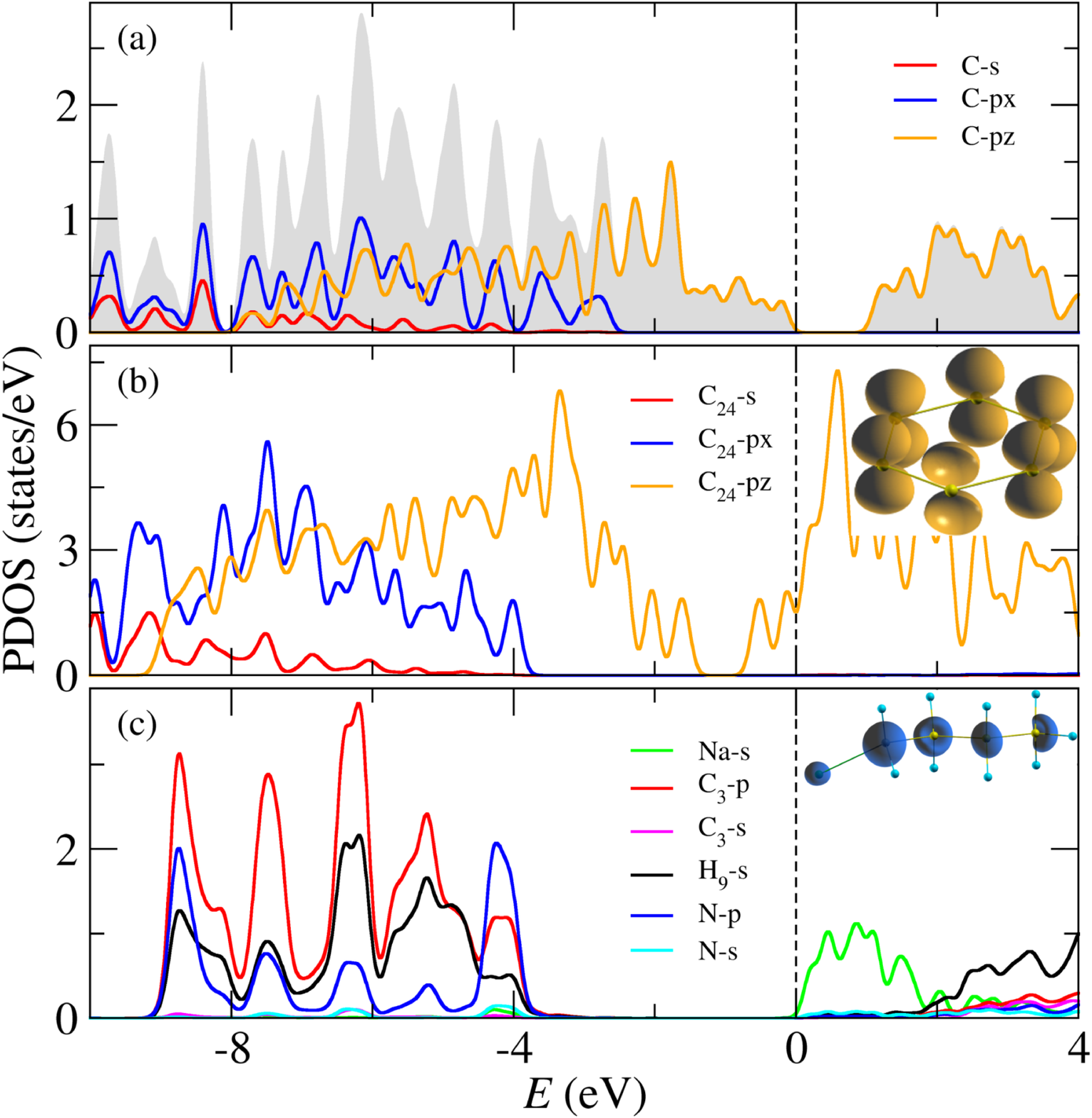} \\
\caption{\label{fig:pdos}(Color online) Partial density of states (PDOS) for graphite and Na-nC3-GIC, calculated with PBEsol XC functional; (a) for graphite with TDOS in grey-colored background, (b) for graphene layer of Na-nC3-GIC, and (c) for Na and other atoms included in propyl amine of Na-nC3-GIC. The dashed vertical line indicates the Fermi level set to be 0 eV.}
\end{center}
\end{figure}

In Figure~\ref{fig:pdos}, we show the partial density of states (PDOS) for Na-nC3-GIC and graphite for comparison. As shown in Figure~\ref{fig:pdos}(a), four valence electrons ($s, p_x, p_y,$ and $p_z$) of carbon atom in graphite make three $sp^2$ hybrid orbitals and one $p_z$ orbital. Here, $sp^2$ hybrid orbitals form strong $\sigma$ bond, while $p_z$ orbitals form weak $\pi$ bond perpendicular to the plane, such that $\pi$ electrons are released easily to take part in the in-plane conduction. The calculated PDOS of Na-nC3-GIC shown in Figure~\ref{fig:pdos}(b)-(c) indicates that Na atom releases its valence $s$ electron, becoming cation (i.e., electron donor), which occupies anti-bonding $\pi^*$ $(p_z)$ orbitals of carbon atoms of graphene layer (i.e., electron acceptor). We can find the hybridization between $\sigma$ bonding orbitals of carbon atoms and molecular orbitals of alkylamine molecule, indicating strong interaction between the electrons. Inset figures are the calculated integrated local density of orbitals 4 eV below or above the Fermi level, showing clearly $\pi$ bonding characteristics of carbon hexagon in graphene layer, and molecular orbitals of cointercalants. The tendency is similar in other AM-nC$x$-GICs.

To obtain more intuitive insights into the charge transferring in AM-nC$x$-GICs, the electronic density difference ($\Delta\rho$) is calculated as the difference between the electronic density of the AM-nC$x$-GICs ($\rho_{\text{M(nC}x\text{)C}_{24}}$) and those of the graphene layers ($\rho_{\text{C}_{24}}$), alkali atom ($\rho_{\text{M}}$) and alkylamine ($\rho_{\text{nC}x}$) as follows,
\begin{equation}
\label{eq:deltarho}
\Delta\rho=\rho_{\text{M(nC}x\text{)C}_{24}}-(\rho_{\text{M}}+\rho_{\text{nC}x}+\rho_{\text{C}_{24}}).
\end{equation}
\begin{figure}[b!]
\begin{center}
\includegraphics[clip=true,scale=0.21]{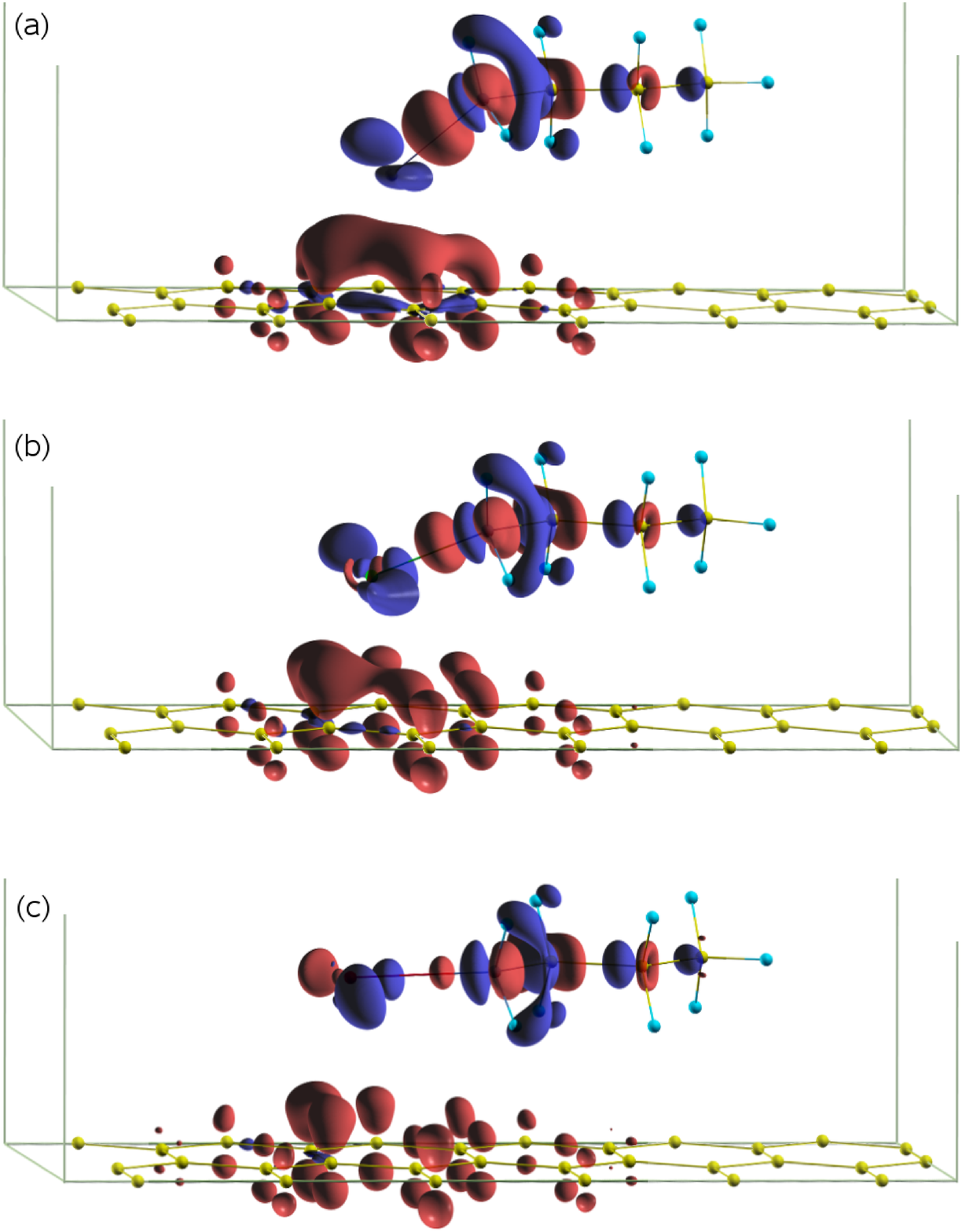}
\caption{\label{fig:dendiff}(Color online) Electronic density difference between electronic density of AM-nC3-GICs and those of the graphene layer, alkali atom and alkylamine; (a) for Li, (b) for Na, and (c) for K. Isosurface figures are evaluated at $\pm$0.003 $\vert e\vert$/\AA$^3$, where red-colored positive value (blue-colored negative value) represents the electronic density accumulation (loss).}
\end{center}
\end{figure}
Figure~\ref{fig:dendiff} shows the electronic density accumulation (red color) and loss (blue color) in AM-nC3-GICs as typical examples. We can see clearly in the figures that the valence electrons of alkali atoms has transferred almost totally to the bonding carbon rings in graphene layer, while in alkylamine molecules the give-and-take of electrons is occurred in intramolecular way or by the result of interaction with other atoms.

Finally, we have calculated the atomic populations on each atom by using Mulliken, Hirshfeld and Voronoi method to quantitatively estimate the charge transfer. For the sake of clarity, we assessed the total populations of alkylamine molecule and of graphene layer by summation of atomic populations of constituent elements. The results obtained by using SIESTA code are presented in Table.~\ref{tab:charge}. Here, positive (negative) value denotes the deficiency (excess) of electrons in the atom or group in the unit of $|e|$. We can find the substantial identity with negligible differences between the three methods. No matter what method is adopted, all the alkali atoms lose electrons while graphene layers gain those and alkylamine molecules lose or gain a little electron depending on the alkyl groups or even considered methods. The extent of electron loss from alkali atoms and thus of electron gain by carbon rings increases going from Li to K, indicating more ionic bond and thus stronger binding. Such atomic population analysis is in good agreement with the structural and energetic analysis discussed above.
\begin{table*}[!th]
\small
\begin{center}
\caption{\label{tab:charge}Atomic charge populations of alkali atoms (AM: Li, Na, and K), alkylamine molecules (nC$x$: x=1 (methyl), 2 (ethyl), 3 (propyl), and 4 (butyl)), and graphene layer (C$_{24}$), calculated by using Mulliken, Hirshfeld, and Voronoi methods. Positive (negative) value presents deficiency (excess) of electrons in the atom or group in the unit of $|e|$.}
\begin{tabular}{lcccc|ccc|ccc}
\hline
      &       & \multicolumn{3}{c|}{Mulliken} & \multicolumn{3}{c|}{Hirshfeld} & \multicolumn{3}{c}{Voronoi} \\
\cline{3-11} 
Alkyl & Metal & AM & nC$x$ & C$_{24}$ & AM & nC$x$ & C$_{24}$ & AM & nC$x$ & C$_{24}$ \\
\hline
       & Li   & 0.190 &  0.140 & -0.330 & 0.243 &  0.096 & -0.337 & 0.192 & 0.176 & -0.367 \\
Methyl & Na   & 0.347 &  0.056 & -0.402 & 0.333 &  0.012 & -0.345 & 0.291 & 0.091 & -0.387 \\
       & K    & 0.431 & -0.020 & -0.411 & 0.335 & -0.024 & -0.310 & 0.278 & 0.064 & -0.342 \\
\hline
       & Li   & 0.261 &  0.101 & -0.362 & 0.290 &  0.046 & -0.335 & 0.244 & 0.118 & -0.363 \\
Ethyl  & Na   & 0.364 &  0.035 & -0.399 & 0.345 & -0.031 & -0.315 & 0.299 & 0.069 & -0.365 \\
       & K    & 0.430 & -0.057 & -0.371 & 0.349 & -0.073 & -0.276 & 0.288 & 0.032 & -0.320 \\
\hline
       & Li   & 0.271 &  0.076 & -0.347 & 0.298 &  0.001 & -0.301 & 0.243 &  0.092 & -0.338 \\
Propyl & Na   & 0.369 &  0.028 & -0.396 & 0.320 & -0.011 & -0.314 & 0.258 &  0.089 & -0.349 \\
       & K    & 0.417 & -0.071 & -0.348 & 0.336 & -0.104 & -0.232 & 0.276 &  0.004 & -0.282 \\
\hline
       & Li   & 0.203 &  0.072 & -0.276 & 0.252 & -0.008 & -0.244 & 0.203 &  0.082 & -0.285 \\
Butyl  & Na   & 0.304 & -0.013 & -0.290 & 0.309 & -0.108 & -0.198 & 0.279 & -0.023 & -0.258 \\
       & K    & 0.398 & -0.097 & -0.301 & 0.321 & -0.143 & -0.179 & 0.261 & -0.028 & -0.231 \\
\hline
\end{tabular}
\normalsize
\end{center}
\end{table*}

\section{\label{sec:con}Conclusions}
With {\it ab initio} DFT calculations, we have investigated the structural, energetic, and electronic properties of AM-nC$x$-GICs containing alkali atoms and alkylamine molecules as cointercalants, motivated by the need of development of high performance anode materials for NIBs. The crystalline models of stage-I t-GICs were built using orthorhombic unit cells and the structural properties including the graphene layer separation $d_{\text{C-C}}$ and the distances from alkali atom to graphene layer $d_{\text{M-C}}$ and to nitrogen atom $d_{\text{M-N}}$ were determined by performing variable cell relaxations with different XC functionals such as PBE and PBEsol within GGA, vdW-DF4, and PZ within LDA. It was confirmed that PZ gives the best layer separation for graphite, while PBEsol yields not so bad $d_{\text{C-C}}$ for graphite and moreover the closest value to the experiment for Li-nC3-GIC. We have observed that with the increase in the atomic number of alkali atoms the layer separation $d_{\text{C-C}}$ decreases while $d_{\text{M-C}}$ and $d_{\text{M-N}}$ increases, implying stronger interlayer binding for more electropositive alkali cation.

We have calculated the formation energies, interlayer binding energies and energy barriers for migration of alkali atoms in AM-nC3-GICs using PBEsol functional. It was found that Na- and K-nC$x$-GICs are stable due to their negative formation energies but Li-nC$x$-GICs are unstable. The calculated interlayer binding energies, -0.511 eV, -0.956 eV, and -1.283 eV for Li-, Na-, and K-nC3-GICs, confirm the increase in binding strength going from Li to K. The energy barriers for the migrations were estimated to be 0.556 eV, 0.276 eV, and 0.157 eV for Li-, Na-, and K-nC3-GICs, indicating more smoothly diffuse of larger ionic radius cation in graphite.

We have performed the analysis of DOS, electronic density differences, and atomic charge populations to consider the charge transferring in AM-nC3-GICs. It has become clear from the TDOS and PDOS analysis that all the AM-nC$x$-GICs studied here are metallic and alkali atoms lose their valence $s$ electrons which occupy anti-bonding $\pi^*$ $(p_z)$ orbitals of carbon atoms of graphene layer. This was confirmed by electronic density differences and atomic charge populations as well. With all these findings, we can insist that cointercalation of alkali atoms together with alkylamine molecules into graphite can make it easy to form stage-I sodium GICs that can be used as anode materials for NIBs.

\section*{\label{ack}Acknowledgments}
This work was supported partially from the Committee of Education, Democratic People's Republic of Korea, under the project entitled ``Strong correlation phenomena at superhard, superconducting and nano materials'' (grant number 02-2014). The simulations have been carried out on the HP Blade System c7000 (HP BL460c) that is owned and managed by the Faculty of Materials Science, Kim Il Sung University.

\section*{\label{auth}Author information}
Corresponding Author \\
*Chol-Jun Yu: e-mail, ryongnam14@yahoo.com.

\section*{\label{note}Notes}
The authors declare no competing financial interest.

\bibliography{Reference}

\end{document}